\documentclass[fleqn,11pt]{article}

\usepackage{amssymb,amsmath}

\usepackage{amsfonts,amssymb,cite}
\usepackage{epsf,graphicx}

\def\noi{\noindent}
\def\jnumber#1#2{\thispagestyle{empty} \noi\unitlength=1mm
    	\begin{picture}(178,10)
            \put(177,15){\llap{\large\it Grav. Cosmol. No.\,#1, #2}}
                    \end{picture}}

\newcommand{\Title}[1]{\noi {{\Large\bf #1}}\\[1ex]}

\newcommand{\Author}[2]{\noi{\bf #1}\\[2ex]\noi{\normalsize\it #2}\\}

\newcommand{\Rec}[1]{\noi {\it Received #1} \\}

\newcommand{\Abstract}[1]{\vskip 2mm \begin{center}
        \parbox{16.4cm}{\small\noi #1} \end{center}\medskip}

\newcommand{\foom}[1]{\protect\footnotemark[#1]}

\def\email#1#2{\footnotetext[#1]{e-mail: #2}\addtocounter{footnote}{1}}

\def\nqq{\hspace*{-2em}}
\def\nhq{\hspace*{-0.5em}}

\def\Jl#1#2{#1 {\bf #2},\ }

\def\ApJ#1 {\Jl{Astroph. J.}{#1}}
\def\CQG#1 {\Jl{Class. Quantum Grav.}{#1}}
\def\DAN#1 {\Jl{Dokl. AN SSSR}{#1}}
\def\GC#1 {\Jl{Grav. Cosmol.}{#1}}
\def\GRG#1 {\Jl{Gen. Rel. Grav.}{#1}}
\def\JETF#1 {\Jl{Zh. Eksp. Teor. Fiz.}{#1}}
\def\JETP#1 {\Jl{Sov. Phys. JETP}{#1}}
\def\JHEP#1 {\Jl{JHEP}{#1}}
\def\JMP#1 {\Jl{J. Math. Phys.}{#1}}
\def\NPB#1 {\Jl{Nucl. Phys. B}{#1}}
\def\NP#1 {\Jl{Nucl. Phys.}{#1}}
\def\PLA#1 {\Jl{Phys. Lett. A}{#1}}
\def\PLB#1 {\Jl{Phys. Lett. B}{#1}}
\def\PRD#1 {\Jl{Phys. Rev. D}{#1}}
\def\PRL#1 {\Jl{Phys. Rev. Lett.}{#1}}

\def\al{&\nhq}
\def\lal{&&\nqq {}}

\def\beq{\begin{equation}}
\def\eeq{\end{equation}}
\def\bear{\begin{eqnarray}}
\def\bearr{\begin{eqnarray} \lal}
\def\ear{\end{eqnarray}}
\def\earn{\nonumber \end{eqnarray}}
\def\nn{\nonumber\\ {}}

\def\dst{\displaystyle}
\def\tst{\textstyle}
\def\fracd#1#2{{\dst\frac{#1}{#2}}}
\def\fract#1#2{{\tst\frac{#1}{#2}}}

\def\dim{\mathop{\rm dim}\nolimits}

\def\intl{\int\limits}

\begin{document}
\twocolumn[

\jnumber{4}{2016}

\Title{On stable exponential  solutions in Einstein-Gauss-Bonnet cosmology with zero variation of
 $G$ }

\Author{V. D. Ivashchuk\foom 1}
      { Institute of Gravitation and Cosmology of  RUDN  University,\\
        ul. Miklukho-Maklaya, 6, Moscow 117198, Russia, \\
        Center for Gravitation and Fundamental
        Metrology, VNIIMS, \\ Ozyornaya St., 46, Moscow 119361, Russia }

\Rec{July 18, 2016}

\Abstract
{ A $D$-dimensional  gravitational model with a Gauss-Bonnet term and the cosmological term $\Lambda$ is considered. Assuming diagonal cosmological metrics, we find,  for certain $\Lambda \neq 0$
new examples of solutions with an exponential time dependence of two scale factors, governed by two Hubble-like
parameters $H >0$ and $h < 0$, corresponding to submanifolds of dimensions $m$ and $l$, respectively, with 
$(m,l) = (4,2), (5,2), (5,3), (6,7), (7,5), (7,6)$ and $D = 1 + m + l$. Any of these solutions  describes 
an exponential expansion of ``our'' 3-dimensional factor-space with Hubble parameter $H$
and  zero variation of the effective gravitational constant $G$. We also prove  the stability of these solutions 
in the class of cosmological solutions with diagonal  metrics.}


]

\email 1 {ivashchuk@mail.ru \\ The paper is prepared within the RUDN-University program 5-100.}

{ 
 \newcommand{\R}{ {\mathbb R} }

\section{Introduction}

In this paper we consider $D$-dimensional gravitational model
with Gauss-Bonnet term and cosmological term. We note that at present
the so-called Einstein-Gauss-Bonnet (EGB) gravitational model and  its modifications, 
see   \cite{Ishihara}-\cite{Ivas-16} and refs. therein,  are intensively studied in  cosmology,
e.g. for possible explanation  of  accelerating  expansion of the Universe following from
supernovae (type Ia) observational data \cite{Riess,Perl,Kowalski}.

Here we  deal with the cosmological solutions with diagonal metrics  governed by $n >3$ scale factors
depending upon one variable, which is the synchronous time variable. We restrict ourselves by the
solutions with exponential dependence of scale factors and present  six new examples
of such solutions: five - in dimensions $D = 7, 8, 9, 13$ and two - for $D = 14$.
Any of  these solutions  describes   an exponential expansion  of $3$-dimensional factor-space
with Hubble parameters $H > 0$ \cite{Ade}  and has a  constant volume factor of  internal space,
which imply zero variation of the effective gravitational constant $G$ either 
in Jordan or Einstein frame \cite{RZ-98,I-96}, see also  \cite{BIM,IvMel-14} and refs. therein.
These solutions obey the most severe restrictions on  variation of $G$ \cite{Pitjeva}.

We  study   the stability of these solutions in a class of cosmological solutions
with diagonal metrics by using results of  refs. \cite{ErIvKob-16,Ivas-16} and show
 that all solutions, presented here, are stable. It should be noted that  two  exponential solutions
with two factor spaces (one of which is expanding and other one - contracting)
and constant $G$ were found for $D = 22, 28$ and $\Lambda = 0$ in ref.  \cite{IvKob}.
In ref. \cite{ErIvKob-16} it was proved that these solutions are stable.

\section{The cosmological model}

The action reads
\begin{equation}
  S =  \int_{M} d^{D}z \sqrt{|g|} \{ \alpha_1 (R[g] - 2 \Lambda) +
              \alpha_2 {\cal L}_2[g] \},
 \label{1}
\end{equation}
where $g = g_{MN} dz^{M} \otimes dz^{N}$ is the metric defined on
the manifold $M$, ${\dim M} = D$, $|g| = |\det (g_{MN})|$, $\Lambda$ is
cosmological term, $${\cal L}_2 = R_{MNPQ} R^{MNPQ} - 4 R_{MN} R^{MN} +R^2$$
is standard Gauss-Bonnet term and  $\alpha_1$, $\alpha_2$ are
non-zero constants.

Here we consider the manifold
\begin{equation}
   M = \R  \times   M_1 \times \ldots \times M_n \label{2.1}
\end{equation}
with the metric
\begin{equation}
   g= - d t \otimes d t  +
      \sum_{i=1}^{n} B_i e^{2v^i t} dy^i \otimes dy^i,
  \label{2.2}
\end{equation}
where   $B_i > 0$ are arbitrary constants, $i = 1, \dots, n$, and
$M_1, ...,  M_n$  are one-dimensional manifolds (either $\R$ or $S^1$)
and $n > 3$.

Equations of motion for the action (\ref{1}) 
give us the set of  polynomial equations \cite{ErIvKob-16}
\begin{eqnarray}
  G_{ij} v^i v^j + 2 \Lambda
  - \alpha   G_{ijkl} v^i v^j v^k v^l = 0,  \label{2.3} \\
    \left[ 2   G_{ij} v^j
    - \frac{4}{3} \alpha  G_{ijkl}  v^j v^k v^l \right] \sum_{i=1}^n v^i \nn
    - \frac{2}{3}   G_{ij} v^i v^j  +  \frac{8}{3} \Lambda = 0,
   \label{2.4}
\end{eqnarray}
$i = 1,\ldots, n$, where  $\alpha = \alpha_2/\alpha_1$. Here
$G_{ij} = \delta_{ij} -1$ and $G_{ijkl}  = G_{ij} G_{ik} G_{il} G_{jk} G_{jl} G_{kl}$
are, respectively, the components of two  metrics on  $\R^{n}$ \cite{Iv-09,Iv-10}. 
The first one is  2-metric and the second one is a Finslerian 4-metric.
For $n > 3$ we get a set of forth-order polynomial  equations.

For $\Lambda =0$ and $n > 3$ the set of equations (\ref{2.3}) and (\ref{2.4})
has an isotropic solution $v^1 = \ldots = v^n = H$ only if $\alpha  < 0$ \cite{Iv-09,Iv-10}.
This solution was generalized in \cite{ChPavTop} to the case $\Lambda \neq 0$.

It was shown in \cite{Iv-09,Iv-10} that there are no more than
three different  numbers among  $v^1,\dots ,v^n$ when $\Lambda =0$. This is valid also
for  $\Lambda \neq 0$ if $\sum_{i = 1}^{n} v^i \neq 0$  \cite{Ivas-16}.

\section{Solutions with constant $G$}

In this section we present several solutions to the set of equations (\ref{2.3}), (\ref{2.4})
of the following form
\begin{equation}
  \label{3.1}
   v =(H, \ldots, H, h, \ldots, h).
\end{equation}
where $H$ is the ``Hubble-like'' parameter corresponding  to $m$-dimensional subspace
with $m > 3$ and $h$ is the ``Hubble-like'' parameter   corresponding to $l$-dimensional
subspace, $l>1$. We put $H > 0$ for a description of an accelerated expansion of
$3$-dimensional subspace (which may describe our Universe) and also put
\begin{equation}
  \label{3.2}
 h = - (m-3) H/l <  0
\end{equation}
for a  description of a zero variation of the effective gravitational constant $G$.

We remind that  the effective gravitational constant $G = G_{eff}$
in the Brans-Dicke-Jordan (or simply Jordan) frame \cite{RZ-98} (see also \cite{I-96})
is proportional to the inverse volume scale factor
of the internal space, see \cite{BIM,IvMel-14} and refs. therein.

According to ansatz (\ref{3.1}),  the $m$-dimensional subspace is expanding
with the Hubble parameter $H >0$, while $l$-dimensional subspace is contracting
with the ``Hubble-like''  parameter $h <0$.

For  $\Lambda =0$, $m=11$,  $l=16$ and $\alpha = 1$ the  solution with
$H= \frac{1}{\sqrt{15}}$, $h= -\frac{1}{2 \sqrt{15}}$, describing  zero variation of $G$,
was found in  \cite{IvKob}.
Another solution of such type  with $\Lambda =0$, $H=\frac{1}{6}$, $h=-\frac{1}{3}$
and constant $G$ appears for $m=15$, $l=6$ and $\alpha = 1$ \cite{IvKob}.
It was proved in \cite{ErIvKob-16} that these two solutions are stable.

Here we present  three solutions with constant $G$ for $\alpha < 0$:
\begin{equation}
 \label{3.4}
  H = \frac{1}{\sqrt{6 |\alpha|}}, \qquad  h =  - \frac{1}{2\sqrt{6 |\alpha|}}
\end{equation}
    for $\Lambda = 7/(8|\alpha|)$, $(m,l)= (4,2)$;
\begin{equation}
   \label{3.5}
   H = \frac{1}{\sqrt{8 |\alpha|}}, \qquad  h =  - \frac{1}{\sqrt{8 |\alpha|}}
\end{equation}
for $\Lambda = 17/(16|\alpha|)$, $(m,l)= (5,2)$ and
\begin{equation}
    \label{3.6}
   H = \frac{3}{2\sqrt{10 |\alpha|}}, \qquad   h =  - \frac{1}{\sqrt{10 |\alpha|}}
\end{equation}
for $\Lambda = 177/(80 |\alpha|)$, $(m,l)= (5,3)$.

We also present  three solutions with constant $G$ for  $\alpha > 0$:
\begin{equation}
  H = \frac{7}{2\sqrt{5\alpha}}, \qquad  h = -  \frac{3}{2\sqrt{5 \alpha}}
          \label{3.7}
\end{equation}
for $\Lambda =  -177.45 \alpha^{-1}$,  $(m,l)= (6,7)$;
\begin{equation}
 H =  \frac{5}{6 \sqrt{\alpha}}, \qquad  h = - \frac{2}{3 \sqrt{\alpha}},
      \label{3.8}
 \end{equation}
for $\Lambda =  - 155/(6 \alpha)$, $(m,l)= (7,5)$, and
\begin{equation}
    H = \frac{3}{2 \sqrt{5\alpha}}, \qquad h = -  \frac{1}{\sqrt{5 \alpha}}
        \label{3.9}
\end{equation}
for $\Lambda = - 8.7  \alpha^{-1}$,  $(m,l)= (7,6)$. 

All six solutions may be verified by MAPLE. The derivation of a more  general  
class of solutions will be  given in a separate paper.

\section{Stability analysis}

In \cite{ErIvKob-16,Ivas-16} we restricted ourselves by exponential solutions (\ref{2.2})
with non-static volume factor, which is proportional to $\exp(\sum_{i = 1}^{n} v^i t)$,
i.e. we put
\begin{equation}
  K = K(v) = \sum_{i = 1}^{n} v^i \neq 0.
  \label{4.1}
\end{equation}

We  put the following restriction on the matrix 
$L =(L_{ij}(v)) = (2 G_{ij} - 4 \alpha G_{ijks} v^k v^s)$ \cite{ErIvKob-16,Ivas-16}
\begin{equation}
  ({\rm R }) \quad  \det (L_{ij}(v)) \neq 0.
  \label{4.2}
\end{equation}

For general cosmological setup with the metric 
$g= - dt \otimes dt + \sum_{i=1}^{n} e^{2\beta^i(t)}  dy^i \otimes dy^i$,
we obtained in \cite{ErIvKob-16,Ivas-16}
the (mixed) set of algebraic and differential equations 
\begin{eqnarray}
        f_0(h) = 0,    \label{4.5a}   \\
        f_i(\dot{h},h) =0,
        \label{4.6a}
\end{eqnarray}
$i = 1, \dots, n$, where $h = h(t) = (h^i(t)) = (\dot{\beta}_i(t))$ is the set of so-called  ``Hubble-like'' parameters; $f_0(h)$ and $f_i(\dot{h},h)$ are polynomials of the fourth order in $h^i$;
$f_i(\dot{h},h)$ are polynomials of the first order in $\dot{h}^i$, see \cite{ErIvKob-16,Ivas-16}.
The fixed point solutions  $h^i(t) = v^i$ ($i = 1, \dots, n$) to eqs. (\ref{4.5a}), (\ref{4.6a})
correspond to exponential solutions  (\ref{2.2}), which  obey eqs. (\ref{2.3}),  (\ref{2.4}).

It was proved in \cite{Ivas-16} that a fixed point solution
$(h^i(t)) = (v^i)$ ($i = 1, \dots, n$; $n >3$) to eqs. (\ref{4.5a}), (\ref{4.6a})
obeying restrictions (\ref{4.1}), (\ref{4.2}) is  stable under perturbations
$h^i(t) = v^i +  \delta h^i(t)$, $i = 1,\ldots, n$,  (as $t \to + \infty$)  if
\begin{equation}
 K(v) = \sum_{k = 1}^{n} v^k > 0
 \label{4.1a}
\end{equation}
and  it is unstable (as $t \to + \infty$) if $K(v) = \sum_{k = 1}^{n} v^k < 0$.

It was shown in  \cite{Ivas-16} that  for  the vector $v$ from  (\ref{3.1}), obeying
\begin{equation}
  m H + lh \neq 0, \qquad  H \neq h,
  \label{4.4}
\end{equation}
the matrix $L$ has a block-diagonal form
\begin{equation}
 (L_{ij}) = {\rm diag}(L_{\mu \nu}, L_{\alpha \beta} ),
 \label{4.5}
\end{equation}
where
\begin{eqnarray}
  L_{\mu \nu} =  G_{\mu \nu} (2 + 4 \alpha S_{HH}),
  \label{4.6HH}     \\
  L_{\alpha \beta} = G_{\alpha \beta} (2 + 4 \alpha S_{hh}) 
  \label{4.6hh}
\end{eqnarray}
and
\begin{eqnarray}
  S_{HH} =  (m-2)(m -3) H^2  \nonumber  \\
         + 2(m-2)lHh + l(l - 1)h^2 ,
  \label{4.7}   \\
  S_{hh} = m(m-1)H^2 \nonumber \\
          + 2m(l - 2)Hh+ (l- 2)(l- 3)h^2.
  \label{4.8}
\end{eqnarray}

The matrix (\ref{4.5}) is invertible if and only if  $m > 1$,  $l > 1$ and
\begin{equation}
 S_{HH} \neq - \frac{1}{2 \alpha}, \qquad  S_{hh} \neq - \frac{1}{2 \alpha}.
 \label{4.9}
\end{equation}
We remind that the matrices  $(G_{\mu \nu}) = (\delta_{\mu \nu} -1 )$ and
$(G_{\alpha \beta}) = (\delta_{\alpha \beta} - 1)$ are invertible only if  $m > 1$ and $l > 1$.

The first condition (\ref{4.1a}) is obeyed for the solutions under consideration
since due to (\ref{3.2}) we get  $K(v) = 3H > 0$ \cite{Ivas-16}.

Now, let us verify the second condition (\ref{4.9}). The calculations give us
\begin{eqnarray}
  (- 2 \alpha)  S_{HH} =  - 0.5, \ -1, \ -1.5, \ 21, \ 10, \  6 ,
  \label{4.10}   \\
  (- 2 \alpha)  S_{hh} = 4, \ 5, \ 6, \ -39,\ -17, \ - 9
  \label{4.11}
\end{eqnarray}
for  the solutions with $(m,l) = (4,2), (5,2), (5,3), \\ (6,7), (7,5), (7,6)$, respectively.
Thus, conditions  (\ref{4.9}) are satisfied for all our solutions. Hence  all six solutions
are stable in a class of cosmological solutions with diagonal  metrics.

\section{Conclusions}

We have considered the  $D$-dimensional  Einstein-Gauss-Bonnet (EGB) model
with the $\Lambda$-term.  By using the  ansatz with diagonal  cosmological  metrics,
we have found, for certain $\Lambda \neq 0$ and $\alpha = \alpha_2 / \alpha_1 $,
six new solutions with exponential dependence of two scale factors
with respect to synchronous time variable $t$ in dimensions $D = 1 + m + n$,
where  $(m,l) = (4,2), (5,2), (5,3), (6,7), (7,5), (7,6)$.
Here $m > 3$ is the dimension  of the expanding subspace and $l > 1$
is the dimension of  contracting  subspace.
The first three solutions correspond to $\alpha < 0$, while other three solutions
correspond to $\alpha > 0$.

Any of these solutions describes an exponential expansion of ``our'' 3-dimensional factor-space with
the Hubble parameter $H > 0$ and anisotropic behaviour of $(m-3+ l)$-dimensional internal space:
expanding in $(m-3)$ dimensions (with Hubble-like parameter $H$) and contracting in $l$ dimensions
(with Hubble-like parameter $h < 0$).
Each solution   has a constant volume factor of internal space and hence it describes
zero variation of the  effective gravitational constant $G$.  By using results of ref. \cite{Ivas-16}
we  have proved that all these solutions  are stable as  $t \to + \infty$.


 {\bf Acknowledgments}

This paper was funded by the Ministry of Education and Science of the Russian Federation
in the Program to increase the competitiveness of Peoples' Friendship University
(RUDN University) among the world's leading research and education centers in the 2016-2020
and  by the  Russian Foundation for Basic Research,  grant  Nr. 16-02-00602.

\end{document}